\documentclass[twocolumn,pre,a4paper,notitlepage,showpacs]{revtex4-1}
\usepackage{graphicx}
\usepackage{amssymb}
\usepackage{amsmath}
\usepackage[dvipsnames]{xcolor}
\usepackage{colonequals}
\usepackage[utf8]{inputenc}
\usepackage{soul}
\usepackage{cancel}
\newcommand{\kBT}{k_\mathrm{B}T}
\newcommand{\vek}[1]{\mathbf{#1}}
\newcommand{\ctimes}{\!\stackrel{\times}{\,}\!}

\newcommand{\beq}{\begin{eqnarray}}
\newcommand{\eeq}{\end{eqnarray}}
\newcommand{\rb}[1]{\left( #1 \right)}
\newcommand{\kett}[1]{| #1 \rangle\!\rangle }
\newcommand{\braa}[1]{\langle\!\langle #1|}
\newcommand{\eww}[1]{\langle\! \langle #1\rangle\! \rangle}

\begin{document}
\title{On the waiting time distribution for continuous stochastic systems}

\author{Robert Gernert$^1$}
\email{robert.gernert@tu-berlin.de}
\author{Clive Emary$^2$}
\email{c.emary@hull.ac.uk}
\author{Sabine H.~L.~Klapp$^1$}
\email{klapp@physik.tu-berlin.de}
\affiliation{
$^1$ Institut f\"ur Theoretische Physik, Sekr.~EW 7--1,
Technische Universit\"at Berlin, Hardenbergstrasse 36,
D-10623 Berlin, Germany
\\
$^2$ Department of Physics and Mathematics,
University of Hull, Kingston-upon-Hull, United Kingdom%
}
\date{\today}

\begin{abstract}
The waiting time distribution (WTD) is a common tool for analysing discrete stochastic processes in classical and quantum systems. However, there are many physical examples
where the dynamics is continuous and only approximately discrete, or where it is favourable to discuss the dynamics on a discretized and a continuous level in parallel. An example is the hindered motion of particles through potential landscapes with barriers. In the present paper we propose a consistent generalisation of the WTD from the discrete case
to situations where the particles perform continuous barrier-crossing characterised by a finite duration. To this end, we introduce a recipe to calculate the WTD 
from the Fokker-Planck/Smoluchowski equation. 
In contrast to the closely related first passage time distribution (FPTD), which is frequently used to describe continuous processes,
the WTD contains information about the direction of motion. As an application, we consider the paradigmatic example of an overdamped particle diffusing through a washboard potential. 
To verify the approach and to elucidate its numerical implications, we compare the WTD defined via the Smoluchowski equation with data from 
direct simulation of the underlying Langevin equation and find full consistency provided that the jumps in the Langevin approach are defined properly.
Moreover, for sufficiently large energy barriers, the WTD defined via the Smoluchowski equation becomes consistent with that resulting from the analytical solution of a (two-state) master equation model for the short-time dynamics developed previous by us [PRE \textbf{86}, 061135 (2012)]. Thus, our approach ``interpolates'' between these two types of stochastic motion. We illustrate our approach for both symmetric systems and systems under constant force.
\end{abstract}
\maketitle
\date{\today}

\section{Introduction}
\label{intro}
Waiting time distributions (WTD) measure the distribution of delay times between subsequent hopping events (``jumps'') in a discrete stochastic process.
The concept of a WTD has been introduced already half a century ago for the description of diffusion processes on lattices
by continuous time random walks \cite{montroll65,scher73,haus87}.
Since then, the analysis of transport processes via WTDs has become common for a wide range of classical and quantum systems
whose dynamics can be viewed, at least approximately, as a {\it discrete} stochastic process. Recent applications include, e.g.~in the biophysical area,
the opening kinetics of ion channels \cite{higgins09}, the (un-)binding of chemical bonds \cite{qian13}, and the ``turnover'' events of enzyme molecules \cite{saha12,english06}. 
In econophysics, WTDs are used to analyse stock market fluctuations \cite{mainardi00}. A further important field are diffusion processes of atoms, molecules, or colloids on
corrugated surfaces \cite{skaug13,prager06,ganapathy10,hanes12}.
Although the processes considered there are actually continuous (particles on real surfaces are not bound to lattice sites),
the effective barriers characterising rough surfaces are often much larger than the thermal energy; thus
the motion can be approximately viewed as a sequence of jumps. 
A similar situation occurs in supercooled liquids and glasses, 
where the stochastic processes correspond to hopping of particles between cages (formed by other particles) 
\cite{jung05,chaud08}, to glass rearrangements \cite{ahn13} or, on a more abstract level,
to jumps between ``metabasins'' in configuration space \cite{doliwa03,heuer05,rubner08} or ``states'' in protein folding \mbox{\cite{hartmann14,thul07,higgins09}}.
Finally, within the area of quantum transport, typical applications include 
electron transport in quantum-dot nanostructures \cite{brandes08,flindt09,rajabi13} and tunnelling events in dynamical single-electron emitters \cite{albert11}.
%

Despite these wide and crossdisciplinary applications of WTDs there is, for many systems, no unique recipe to actually calculate this quantity.
Indeed, a straightforward definition is only possible for truly discrete stochastic systems where the dynamics is described by a (Markovian) master equation (ME).
Here, the jumps are {\it a~priori} discrete and have (once they occur) {\it zero} duration. Given a sufficiently simple structure of the ME, it is often 
even possible to calculate the WTD analytically (see recent studies in quantum transport \cite{flindt09,brandes08,albert11} and ion channel kinetics \cite{higgins09}). 
For continuous systems, however, the calculation and
even the very definition of a WTD are not straightforward: and ultimately this is tied to
the unavoidable ambiguity in defining a discrete ``jump'' for a continuous evolving system. 

A frequently-used strategy in such situations is to obtain the WTD numerically from an analysis of (particle) trajectories generated either experimentally or by many-particle computer simulations. Indeed, state-of-the-art experiments allow {\it real time} tracking of the motion of colloids on random surfaces \cite{hanes12,evers13}, on solid-liquid interfaces \cite{skaug13},
or during surface growth \cite{ganapathy10}. Similar trajectory analysis has been performed
in computer simulations (such as Molecular or Brownian Dynamics) of glassy or otherwise frustrated systems (see, e.g., \cite{jung05,heuer05}).
For a continuous system with ``quasi-hopping'' motion, typical trajectories reveal the existence of ``traps'' (in real or configuration space), where the trajectory spends most time,
and fast motion (barrier crossing) in between. However, to extract from that a WTD, one necessarily needs to discretise the space (i.e., to define regions which trajectories can enter or leave), 
yielding some ambiguity in the definition of jumps. In particular, depending on these details of the discretisation,
the resulting jumps can have a {\it finite} duration. Moreover, in many practical applications the measured or simulation-generated trajectories render quite noisy WTDs;
extracting smooth results (and relevant time constants) then requires extensive simulation or experimental time. In view of these difficulties it seems desirable to construct a WTD in a yet alternative way, namely 
on the basis of a \textit{noise-averaged} description of the stochastic process. The natural framework is, of course, the Fokker-Planck equation for the probability density.

In this spirit, we propose in the present paper a simple recipe to calculate the WTD from a Fokker-Planck equation.
Our approach is inspired from the calculation \mbox{\cite{risken84}} of the \textit{first passage time distribution} (FPTD) \mbox{\cite{verechtchaguina06,benichou14,haenggi90,metzler04,gang14}}; 
indeed the two quantities are closely related when we assume the dynamics to be
\mbox{\textit{Markovian}} (i.e., time-homogeneous). 
For symmetric situations our WTD becomes, in fact, identical to the FPTD; however it is more general than the FPTD in that
it allows for directional resolution in asymmetric, continuous situations. We note that in the discrete case, there are already various applications of WTDs to anisotropic situations
\cite{shushin08,hartmann14,thul07,higgins09,rajabi13,albert11,brandes08}. Also, FPTDs with directional resolution have been proposed in the mathematical literature (see e.g.~Refs.~\cite{navarro09,sacerdote14}); however, these have been restricted to the special case of no barriers.
\par%
Considering a particular model system we validate our WTD definition by comparing 
with the WTD following from direct simulation of the underlying Langevin equation of motion.
Specifically, we consider the paradigm example of an overdamped (``Brownian'') particle in a periodic ``washboard'' (e.g., sinusoidal) potential,
a generic model  with relevance in many areas such as particle transport on topologically or energetically 
structured surfaces \cite{Tierno10,Dalle11}, the motion of motor proteins \cite{xing05}, particles in optical lattices \cite{Siler10} and optical line traps \cite{Lopez08}, 
biased Josephson junctions \cite{zapata96,shah07,augello10}, and more generally in biophysical processes \cite{bonnet11,xing05,kay07,Ros05}. Moreover, (non-overdamped) variants of our model also have been used to study certain aspects of the dynamics (such as the appearance of different types of orbits) from a mathematical point of view \cite{carinci13,hairer08}.
\par%
In an earlier study \cite{emary12} we have analysed in detail the short-time dynamics of that system, particularly the plateaus appearing in the mean-squared displacement and the corresponding non-Gaussian parameters for large values of the ratio between barrier height and thermal energy. 
We have demonstrated that, in this limit of deep wells, the continuous dynamics described by the Smoluchowski equation (SE)
can be well described by a ME approach involving two states per well \cite{emary12}. One attractive feature of this ME model is that it is analytically solvable; i.e., properties such as cumulants 
of the density distribution can be calculated by minimal effort. 

Here we use this ME approach as a third, {\it analytical} route to calculate the WTD. We show that the WTD obtained from the SE and BD approach
becomes consistent with that of the discrete case in the appropriate limit (deep wells). 
As wells become shallower, the ME approach becomes less good, an indicator being that the duration of the transitions between different wells increases. In summary, we suggest a consistent generalisation of WTD from the discrete, zero-duration jump case 
to the continuous, extended-jump one.  

The rest of this paper is organised as follows. In Sec.~\mbox{\ref{model}} we specify the equations of motion of our model and discuss representative trajectories. The three approaches to the WTD are introduced in Sec.~\mbox{\ref{routeswtd}}, supplemented by two appendices where we present detailed information on the numerical calculation. Numerical results are presented and discussed in Sec.~\mbox{\ref{results}}, and in Sec.~\mbox{\ref{conclusion}} we summarise our findings.

\section{Model}
\label{model}%
We begin by discussing a concrete model as an illustration of the nature of trajectories and jumps in the class of systems we investigate here. We consider a particle in the tilted washboard potential
\begin{equation}
\label{eq:pot}
	u(x)=u_0 \sin^2\frac{\pi x}{a}-Fx,
\end{equation}
where $a$ is the period of the potential, and the parameters $u_0$ and $F$
describe the depth and tilt of the potential, respectively. Under overdamped conditions (strong friction),
the equation of motion of the particle in one dimension is given by
\begin{equation}
	\gamma\dot{x}(t)=-u'(x)+\sqrt{2\gamma k_{\mathrm{B}}T}\,\xi(t)
	\label{LE}
	.
\end{equation}
The thermal energy is denoted by $\kBT$, $\gamma=k_{\mathrm{B}}T/D_0$ is the friction constant (involving the diffusion constant
$D_0$ of the free system), $-u'(x)=-du/dx$ is the force, and $\xi(t)$ represents Gaussian white noise
with properties $\langle \xi(t)\rangle=0$ and $\langle \xi(t)\xi(t')\rangle=\delta(t-t')$. 
\begin{figure}
	\includegraphics[width=\linewidth]{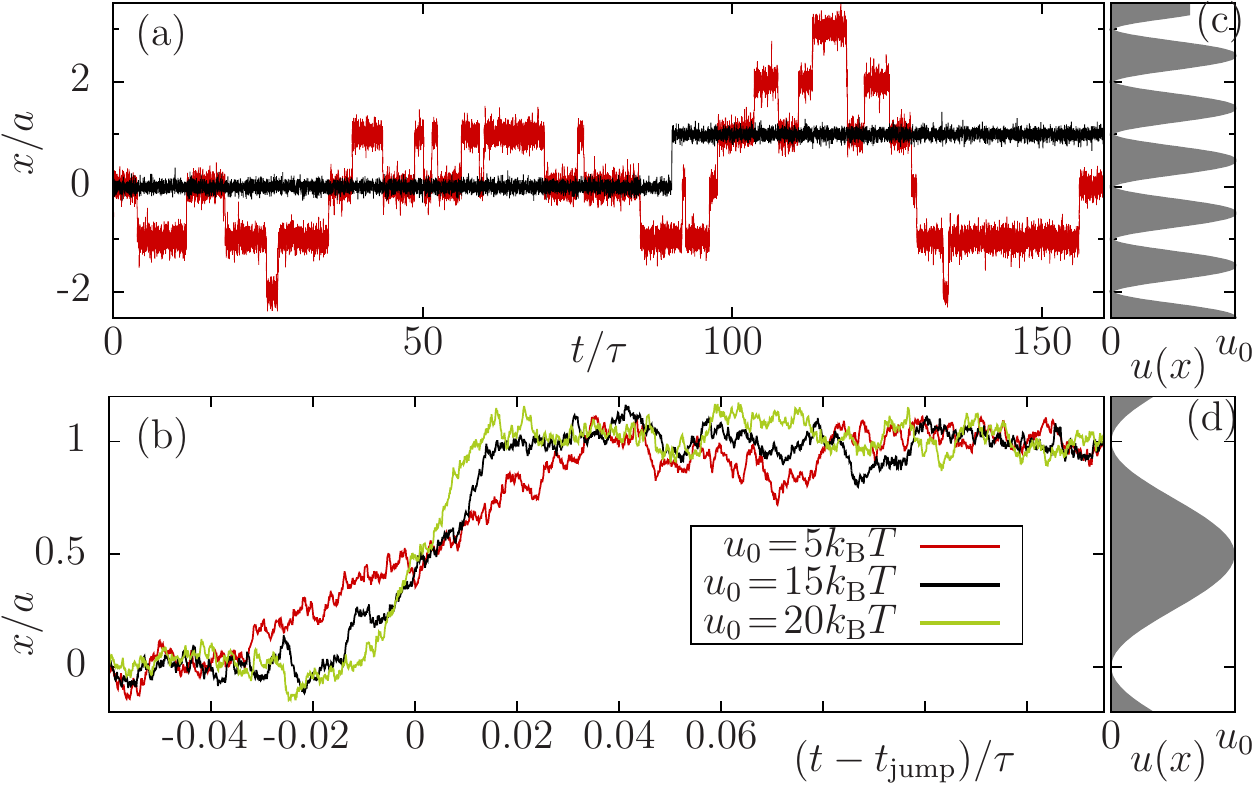}
	\caption{(Color online) (a) Representative trajectories for $F\!=\!0$ for different values of $u_0/k_{\mathrm{B}}T$. (b) To ease the comparison of the jumps, the trajectories are plotted as functions of the relative time $t-t_\mathrm{jump}$ where $t_\mathrm{jump}$ is the arithmetic mean between leaving a potential minimum and arriving at a neighbouring one. We use the time unit $\tau=a^2/D_0$.
	Parts (c) and (d) are sketches of the potential Eq.~(\ref{eq:pot}) corresponding to parts (a) and (b), respectively. The position $x$ is measured in units of the period $a$.}
	\label{trajectories}
\end{figure}%
A comprehensive overview of the different regimes of Eq.~\eqref{LE} in terms of the ratio between noise and external potential or driving force in given in Ref.~\cite{risken84}.
In the present study we focus on cases where the form of the potential is such that it can be meaningfully divided in a number of wells (defined between subsequent maxima). 
This occurs when the difference between
the potential minima and maxima, $\Delta u(u_0,F)$, is large compared to the
thermal energy, i.e., $u_0\gg k_{\mathrm{B}}T$, and when the driving force $F$ is smaller than the so-called critical force $F_\mathrm{c}=u_0\pi/a$, related to the diffusion maximum
\mbox{\cite{Reimann01}} (for $F>F_\mathrm{c}$ the potential minima vanish).

In order to illustrate the motion in such situations we
consider representative particle trajectories. A general discussion of the solutions of the stochastic differential equation \eqref{LE} in terms of their existence and uniqueness is given in Ref.~\cite{kloeden92}. Here we obtain solutions numerically by using 
a Brownian Dynamics (BD) algorithm \cite{ermak75}, i.e.~the Euler-Maruyama scheme \mbox{\cite{kloeden92}}.
\par
\begin{figure}
	\centering
	\includegraphics[width=\linewidth]{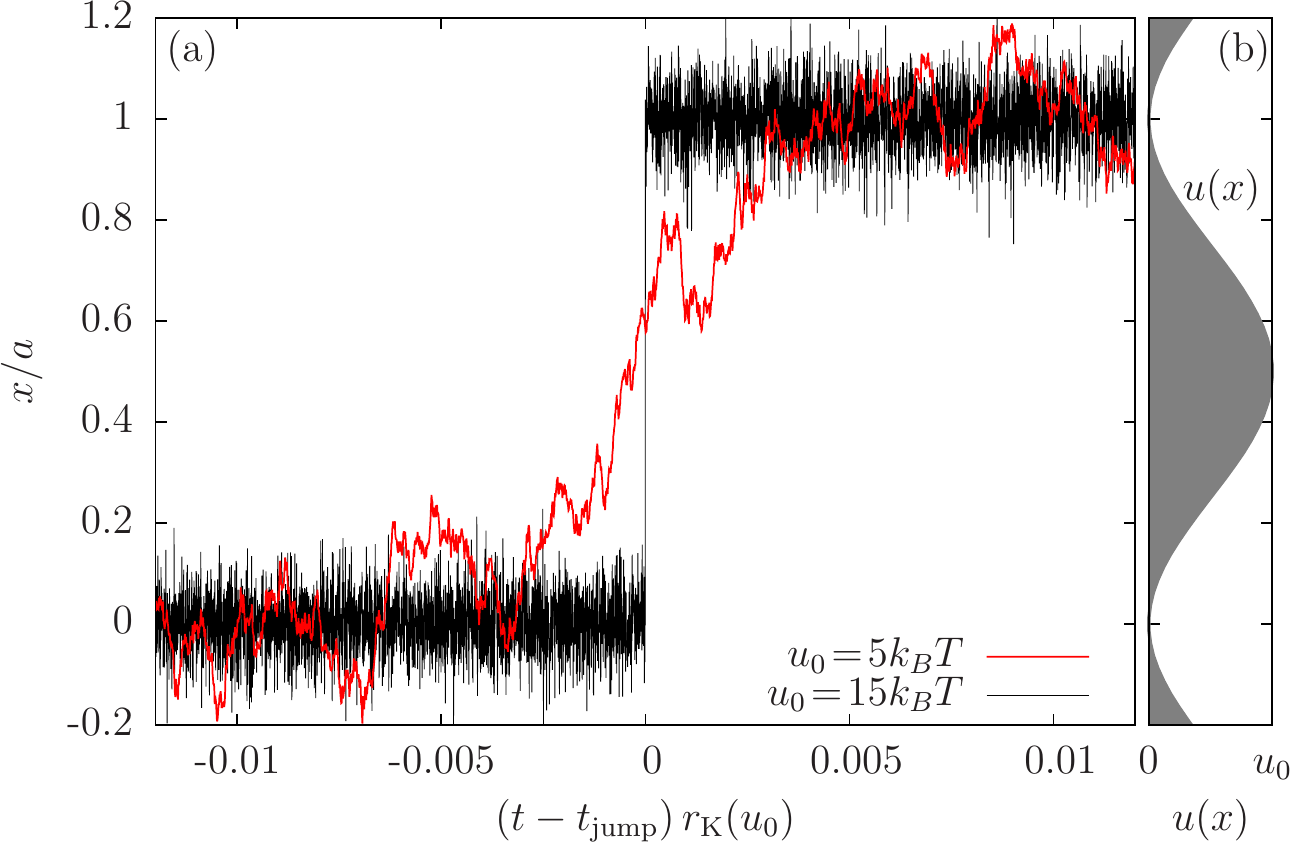}
	\caption{
	  (Color online) Same as Fig.~\ref{trajectories}(b), but with the time axis in part (b) being rescaled by Kramers' rate $r_\text{K}(u_0)$(see main text). With this rescaling, the mean time that the particles spend trapped around a minimum is the same, independent of barrier height.  Relative to this time scale, the jumps in the $u_0=15\kBT$ case appear instantaneous, whereas those for $u_0=5\kBT$ maintain a continuous character.
	}
	\label{trajectories-kramers}
\end{figure}

%
Examplary trajectories for two different values of $u_0/\kBT$ and $F\!=\!0$ are shown in Fig.~\mbox{\ref{trajectories}(a)}. The trajectories in Fig.~\mbox{\ref{trajectories}(a)} clearly reflect that the particle is ``trapped'' for certain times in the regions around the potential minima $x/a \in \mathbb{Z}$. To facilitate the comparison between the trajectories corresponding to different $u_0$ they are plotted in Fig.~\mbox{\ref{trajectories}(b)} as functions of the relative time.  This latter is defined as the difference between the actual time and the time $t_\mathrm{jump}$, which is the arithmetic mean of the times related to the beginning and end of a jump. Here, ``beginning'' and ``end'' refer to leaving of a potential minimum (without going back) and arriving at a different minimum (this definition of a jump will later be called minimum-based definition). 
From Fig.~\mbox{\ref{trajectories}(b)} we see that even for the largest barrier considered, the particle needs a {\it finite} time to cross the barrier. Thus, the motion is still not perfectly discrete.
However, as we may note already here, a different picture on the degree of ``discreteness'' arises when we rescale the time axis with the corresponding Kramers' rate $r_\text{K}(u_0)=D_0/(2\pi \kBT)\sqrt{-u''(x_\mathrm{min}) u''(x_\mathrm{max})}\,\exp(-\Delta u/(\kBT))$ \mbox{\cite{risken84,emary12,kramers40}}, see Fig.~\mbox{\ref{trajectories-kramers}}.
By multiplying the time with this well-known quantity, we take into account that the increase of the potential height alone already leads to an increase of the escape time \mbox{\cite{reimann99}}.  As a consequence the intervals between jumps in the rescaled time become independent of $u_0$ in the long time limit.
Indeed, after rescaling, the data for $u_0=15\kBT$ reflects an essentially discrete motion. However, at $u_0=5\kBT$ we still observe a finite duration of the jump. This clearly demonstrates that the case $u_0=5\kBT$ is at the boundary between discrete and continuous motion.
\section{Routes to calculate the WTD}\label{routeswtd}
\subsection{Direct evaluation via BD simulations}
\label{sec:BD}%
To define the WTD we first need to consider the possible types of jumps. To this end, we note that the stochastic process $x(t)$ [see Eq.~\mbox{\eqref{LE}}] is a Markov process. Thus, the jump characteristics are independent of the history. Further, because the motion is one-dimensional, there are only two directions in which the particle can jump. Finally, Eq.~{\protect\eqref{LE}} obeys the translational symmetry $x\to x\pm a$. Because of these three properties the jumps can be grouped into only two types, namely ``to the right'' and ``to the left'', independent of when or where the jump began.
We label these jump types by the index ``J''. The WTD $w^J(t)$ is then the probability density for the time between an arbitrary jump and a subsequent jump of type $J$. 
\par%
In the context of BD simulations [i.e., direct evaluation of Eq.~\mbox{\eqref{LE}}], the WTD $w^J(t)$ is extracted from a histogram of waiting times extracted from the trajectories pertaining to a given realisation of noise. We note that the WTDs $w^J(t)$ must fulfil the normalisation condition
\begin{gather}
	1=\int_0^\infty\mathrm{d}t\,\sum_J w^J(t)
	,
	\label{normalisation}
\end{gather}
which expresses the fact that the particle leaves its minimum for certain. In terms of the survival probability
\begin{gather}
	S(t)=\int_t^\infty\mathrm{d}t'\,\sum_J w^J(t')
	\label{survivalprob}
\end{gather}
the normalisation is expressed as $S(0)=1$. The sum runs over all jumps types that leave a potential minimum, in our case ``to the right'' and ``to the left''.
\begin{figure}
	\centering
	\includegraphics[width=\linewidth]{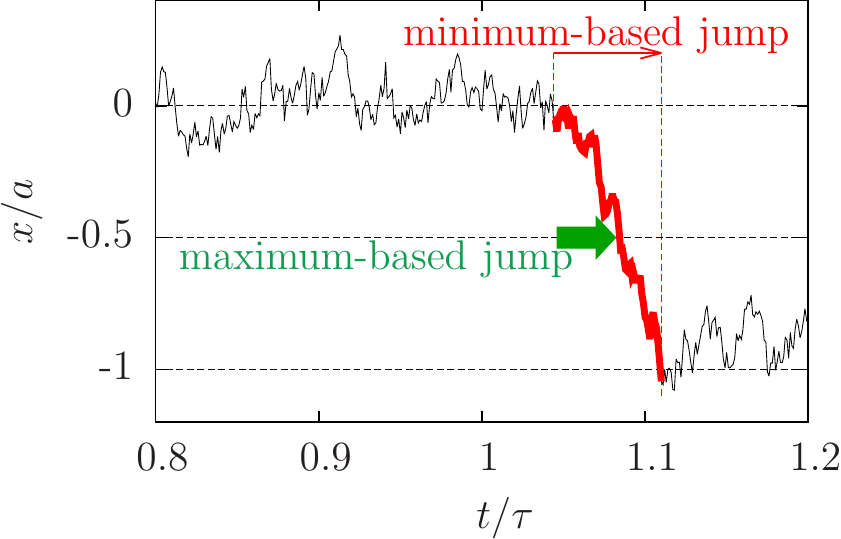}
	\caption{(Color online) 
	Illustration of different `measurement-based jump'
	definitions using a typical trajectory. The thick arrow
	indicates the time at which an instantaneous jump occurs according to the maximum-based definition, i.e.~when the particle crosses the potential maximum between the wells. The thick curve indicates the motion during the jump according to the minimum-based definition, where the jump (having a finite duration) begins with the particle's last departure from a potential minimum (in this example $x\!=\!0$) and ends upon arrival at a neighbouring minimum (here $x\!=\!-a$). The parameters are $u_0\!=\!5\kBT, F\!=\!0$.}
	\label{jumpdef}
\end{figure}%
\par%
In our BD calculations of $w^J(t)$ we consider two definitions of a jump which are illustrated in Fig.~\ref{jumpdef}. Within the first definition, a jump occurs if the particle crosses a potential maximum (\textit{maximum-based definition}). Note that this automatically defines the jump to be instantaneous.
\par%
Within the second definition, a jump involves the entire motion between two neighbouring minima of the potential (\textit{minimum-based definition}). As a consequence, the jump has a finite duration defined as the time between leaving a minimum and arriving at a neighbouring one. 
We call the corresponding probability distribution the jump duration distribution.
Further, we define the waiting time as the time between the end times of two subsequent jumps.
\par%
Numerical details of the calculation are given in Appendix \ref{app:BD}.
\subsubsection{Discussion of the jump definitions}
\begin{figure}
	\centering
	\includegraphics[width=\linewidth]{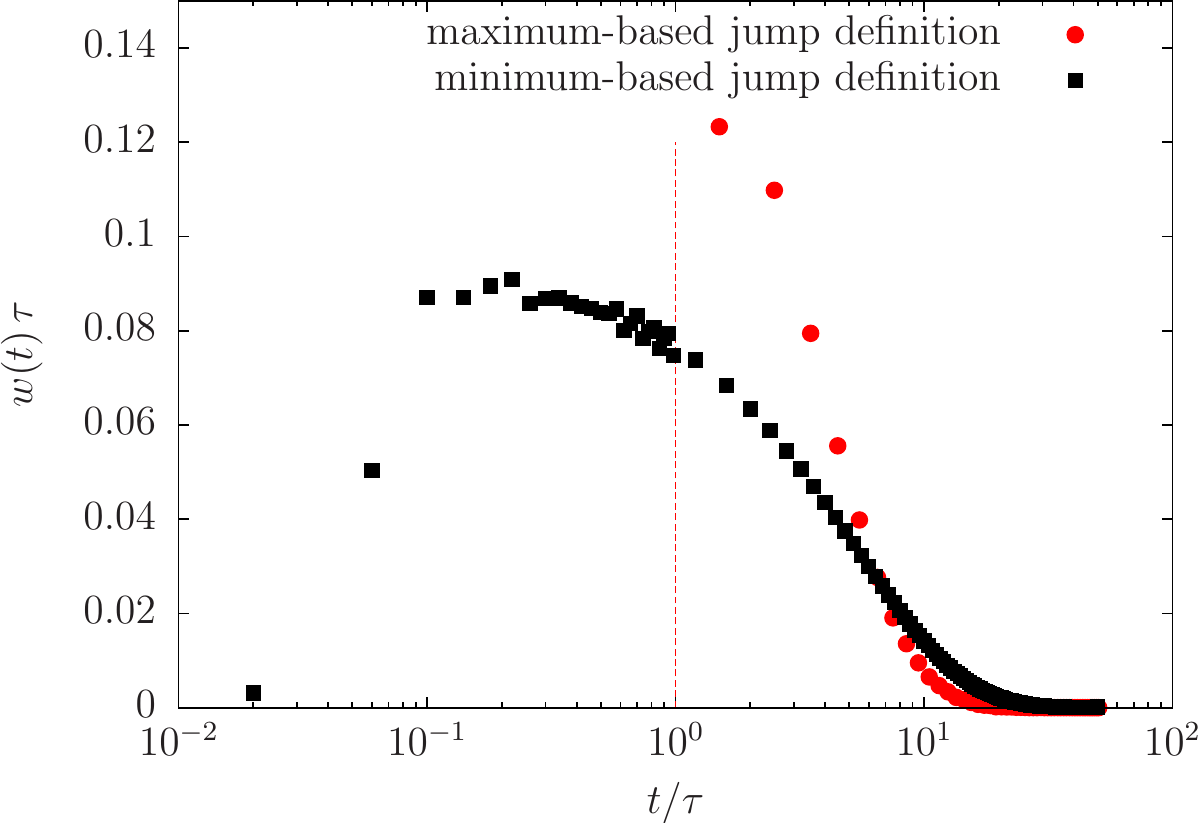}
	\caption{
	  (Color online) Comparison of the WTDs obtained with the two jump definitions described in Sec.~\ref{sec:BD} for $F\!=\!0$ and $u_0\!=\!5\kBT$ (circles and squares). Time is plotted logarithmically. The maximum-based definition breaks down at short waiting times and we suppress these results by introducing a minimal waiting time (``dead time''), here at $t = \tau$ (dashed vertical line).
	}
	\label{wtdjumpdef}
\end{figure}

To investigate the role of the two jump definitions we consider the resulting WTDs for the symmetric case \mbox{$F\!=\!0$}, \mbox{$u_0\!=\!5\kBT$}, plotted in Fig.~\ref{wtdjumpdef}. Because of the symmetry of the potential at $F\!=\!0$, there is only one relevant WTD. Clearly, the choice of the jump definition has a strong impact on the WTD.
In particular, we find that only the minimum-based definition yields a smooth shape of $w^J(t)$.
\par%
On the contrary, the maximum-based jump definition poses several problems. First, during the analysis of the trajectories we noticed that the particle crosses the maximum many times subsequently which generates a lot of spurious (maximum-based) jumps. We deal with this problem by introducing a ``dead time'' after recording a jump which must pass until the next jump can be recorded. We set the ``dead time'' to $1\tau$ (i.e., the time the particle needs to diffuse the distance $\sqrt{\tau/D_0}=a$) to ensure that it leaves the maximum during that time. As a consequence of introducing the dead time, the corresponding WTD is defined to be zero for $t<1\tau$.
\par%
Further, we see in Fig.~\ref{wtdjumpdef} that by using the maximum-based jump definition, waiting times in the range $1\tau \lesssim t\lesssim 5\tau$ are more likely than by using the minimum-based jump definition. The reason is that the particle, once it has gone uphill, is counted to have made a jump in the maximum-based jump definition. On the contrary, the same particle is only counted in the minimum-based jump definition if it goes downhill \emph{in the same direction} (and not backwards). Thus, the maximum-based jump definition frequently records jumps although the particle does not change the potential well. This contradicts our idea of a jump to cross the potential barrier. This problem could be solved by keeping track of the direction in which the particle crosses the maximum. However, the first problem remains (i.e.~accessing short waiting times). We therefore choose the minimum-based jump definition for the further investigation.
\subsection{Definition from the Smoluchowski equation}\label{sec:SE}
The Fokker-Planck equation corresponding to Eq.~\eqref{LE} is given by the Smoluchowski equation (SE)
\begin{align}
	\partial_t p(x,t)&=D_0\frac{\partial^2}{\partial x^2} p(x,t)+\frac{D_0}{\kBT}\frac{\partial}{\partial x}\Big(p(x,t) u'(x)\Big)
	\,,
	\label{FP}
\end{align}
where $p(x,t)\Delta x$ is the probability that the particle is in the interval $[x,x+\Delta x]$ at time $t$.
\par%
Again, we are interested in the distribution of waiting times $t$.
In Sec.~\mbox{\ref{sec:BD}} we defined $w^J(t)$ as the WTD for the jump $J$. Here we associate $J$ with the jumps from an initial position $B$ to a target position $C$. Typically $B$ and $C$ are neighbouring minima of the potential. In order to realise the requirement that the WTD describes the time between \textit{subsequent} jumps we need to exclude processes that lead the particle towards the other neighbouring minimum, which we call $A$.
Correspondingly, we describe the WTD by the function $w_A^{B\to C}(t)$ where $B\to C$ stands for $J$ and the subscript $A$ indicates the excluded position. The definition is as follows:
\par%
\begin{samepage}%
\noindent%
\textbf{Definition:} The quantity $w_{A}^{B\to C}(t)\, \Delta t$, where $\Delta t$ is a small time interval, is the probability that
\begin{subequations}%
\begin{align}
	&\text{\begin{minipage}{0.8\linewidth}\begin{itemize}\item the particle was \emph{for certain} at position $B$ at $t\!=\!0$,\!\!\end{itemize}\end{minipage}}
	\label{def-ic}
	\\
	&\text{\begin{minipage}{0.8\linewidth}\begin{itemize}\item the particle reaches position $C$ in the interval $[t,t+\Delta t]$, and\end{itemize}\end{minipage}}
	\label{def-target}
	\\
	&\text{\begin{minipage}{0.8\linewidth}\begin{itemize}\item the particle neither arrived at $C$ nor $A$ within the time interval $[0,t]$.\end{itemize}\end{minipage}}
	\label{def-stillin}
\end{align}%
\end{subequations}%
\end{samepage}%
Note that we do not care about what happened at earlier times $t<0$, that is, whether the particle moved to $B$ via a jump process or whether it was placed there ``by hand''.
This crucial assumption reflects that we are working in a \emph{Markovian} situation, where the motion from $B$ to $C$ is independent of the history.
\par%
To formulate a recipe to calculate the WTD, we recall that the SE, Eq.~(\ref{FP}), can also be written as a continuity equation, that is,
\begin{align}
	\partial_t p(x,t)+\partial_x j(x,t) &=0
	\label{continuity}
	,
\end{align}
where $j(x,t)=-D_0\left(\partial_{x} p(x,t)+p(x,t) u'(x)/(\kBT)\right)$ is the current density. Equation~(\ref{continuity}) further shows that 
$j(x,t)\,\Delta t$ (or $-j(x,t)\,\Delta t$, respectively) can be interpreted as the probability that the particle crosses position $x$ into positive (negative) $x$-direction during the time interval $[t,t+\Delta t]$.
Consequently $\vert j(C,t)\Delta t\vert$ is the probability that the particle crosses position $C$ in forward or backward direction.
This probability is similar in character to that mentioned in condition \mbox{\eqref{def-target}}. However, $\vert j(C,t)\,\Delta t\vert$ includes the probability that the particle went across $C$ (or $A$), came back, and crosses $C$ again. Such a process violates condition \mbox{\eqref{def-stillin}}.
\par%
Based on these considerations, we propose the following recipe to calculate the WTD $w_{A}^{B\to C}(t)$: 
We initialise the probability density as
 \begin{align}
	p(x,t=0)&=\delta(x-B)
	\label{initialcondition}
\end{align}
according to condition \mbox{\eqref{def-ic}}.
Then, we let $p(x,t)$ evolve in time according to Eq.~(\ref{FP}). However, during this time evolution
we exclude processes from the ensemble of realisations (i.e.~possible trajectories) where  the particle has already reached position $C$ or $A$ [see condition \mbox{\eqref{def-stillin}}].
Therefore $p(x,t)$, which is an average over this reduced set of realisations, fulfils the boundary conditions
\begin{subequations}
	\label{boundarycondition}
	\begin{align}
		p(x,t)&=0,\,x\geq \max(A,C)\quad\forall\, t
		\\
		p(x,t)&=0,\,x\leq \min(A,C)\quad\forall\, t	
		\,.
	\end{align}
\end{subequations}
Equation~(\ref{boundarycondition}) expresses what is often called ``absorbing wall boundary conditions''. Clearly, there is some arbitrariness in defining a suitable position $A$ (the boundary ``on the other side'') for a continuous system in general. However, given that we consider a continuous potential with well-defined wells (as it is the case here) one can give a clear physical meaning to the positions $A$, $B$, and $C$.
\par%
To summarise, we calculate the WTD from the relation
\begin{gather}
	w_{A}^{B\to C}(t)=\vert j^B_{A,C}(C,t)\vert
	\label{wtdSE}
	\,,
\end{gather}
where the notation $j^B_{A,C}$ expresses the dependency of the current density on the initial and boundary conditions given in Eqs.~\eqref{initialcondition} and \eqref{boundarycondition}, respectively.
By using the absolute value in Eq.~\eqref{wtdSE} we take into account the fact that the sign of $j$ depends on the direction of motion. Indeed, motion into the positive (negative) x-direction
implies a positive (negative) sign of $j$. The numerical calculation of the WTD via the SE route is described in appendix B.

Finally, we note that the WTDs defined according to Eq.~\eqref{wtdSE} also fulfil the normalisation condition Eq.~\eqref{normalisation}, that is
\begin{gather}
	\int_0^\infty\mathrm{d}t\left(w_A^{B\to C}(t)+w^{B\to A}_C(t)\right)=1
	\label{normSE}
	.
\end{gather}
%
By definition \cite{risken84}, the first passage time distribution (FPTD) is the probability distribution for the time $t$ the particle needs to leave a given domain for the first time. This definition implicitly assumes that the spatial probability distribution $q(x,\tau)$ to find the particle at position $x$ at an earlier time $\tau<0 \le t$ is known. In the present context, where we are considering a Markovian system, the precise value of $\tau$ is irrelevant. We thus consider the FPTD $F_D(t;q)$ where the domain $D$, in an one-dimensional system, is given by an interval, e.g.~$D=[A,C]$. Further, if the particle was for certain at a position $B$ at $t=0$, we can specialise $q(x)=q^0_B(x)\colonequals\delta(x-B)$. The relation between the FPTD and the WTDs defined before is then given by
\begin{gather}
	F_{[A,C]}(t;q^0_B)=w_{A}^{B\to C}(t)+w_{C}^{B\to A}(t)
	\label{fptd}
	.
\end{gather}
Equation \eqref{fptd} reflects the fact that the FPTD does not contain information about the direction in which the particle left the domain, whereas the WTD does.
\subsection{Master equation}\label{master}

In Ref.~\cite{emary12}, we introduced a simple master equation model of this system.  
In this model, the location of the particle is described by two discrete indices: an index $n$ that describes which of the potential wells the particle finds itself in, and an internal index $\alpha = L,R$ that corresponds to whether the particle is localised in the left or right side of the well.
Relative to the centre of the well at  $x=0$, the loci of these localised states are denoted $x_L$ and $x_R$.
Transitions between these states are assumed to occur at rates  $\gamma^\pm$ to describe hopping within a single well, and $\Gamma^\pm$ to describe hopping between the wells. In both cases, superscript $+$ indicates movement to the right and $-$, movement to the left. 
We choose the inter-well rates, $\Gamma^\pm$, to be twice the corresponding Kramers' rate \cite{emary12,risken84,kramers4thorder}; the intra-well rates, $\gamma^\pm$, as well as the two positions, $x_{L,R}$, are set by fitting the short-time behaviour of the the first two position cumulants to the behaviour of the SE. See Ref.~\cite{emary12} for details.

With the probabilities $p_\alpha^{(n)}$ that the particle is in state $\alpha=L,R$ of the $n$th well arranged into the vector
\beq
  \kett{\rho^{(n)}}
  \equiv
  \rb{
    \begin{array}{c}
      p^{(n)}_L \\
      p^{(n)}_R 
    \end{array}
  }
  ,
\eeq
the system can be described with the master equation
\beq
  \!\!\!
  \!\!\!
  \!\!\!
  \frac{d}{dt} \kett{\rho^{(n)}}
  =
 \mathcal{W}_0\kett{\rho^{(n)}} 
 +
 \mathcal{W}_+
 \kett{\rho^{(n-1)}}
 +
 \mathcal{W}_-
 \kett{\rho^{(n+1)}}
 \label{ME2}
 ,
\eeq
where matrices
\beq
  \mathcal{W}_+ 
  = 
  \rb{
    \begin{array}{cc}
      0 & \Gamma^+ \\
      0 & 0
    \end{array}
  }
  ~~\text{and}~~
  \mathcal{W}_- 
  =
  \rb{
    \begin{array}{cc}
      0 & 0 \\
      \Gamma^- & 0
    \end{array}
  }
\eeq
describe jump processes between the wells and
\beq
  \mathcal{W}_0 
  &=&  
  \rb{
    \begin{array}{cc}
      -\gamma^+ -\Gamma^-& \gamma^- \\
      \gamma^+ & -\gamma^- -\Gamma^+
    \end{array}
  }
\eeq
describes intra-well processes and normalisation.

The waiting time distribution(s) in the ME approach are easy to define but before we do so, we need some more notation.  We define the `expectation value'
$
  \eww{\mathcal{A}}
  \equiv
  \eww{ 1 | \mathcal{A} | \rho_0}
$
where trace and intra-well stationary vectors are defined by
\beq
  \braa{1}\rb{\mathcal{W}_0 + \mathcal{W}_+ + \mathcal{W}_-}
  &=&
  0
  \nonumber\\
  \rb{\mathcal{W}_0 + \mathcal{W}_+ + \mathcal{W}_-}\kett{\rho_0}
  &=&
  0
  .
\eeq
Following \cite{brandes08}, we then obtain the WTDs as
\beq
  w_{ss'}(\tau) \equiv 
  \frac{
    \eww{\mathcal{W}_{s}e^{\mathcal{W}_0 \tau}\mathcal{W}_{s'}}
  }{
    \eww{\mathcal{W}_{s'}}
  }
  ;
  \quad s,s' = \pm
  \label{wtdME}
\eeq
There are four different waiting times defined here, corresponding to the four different combinations of jumps: $s,s' = \pm$.  
In comparing with the results from BD/SE, we will only consider the diagonal WTDs $w_{ss}(\tau)$ as these are found to be the closest analogues of the definitions used in the continuous system.  The reason for this is that in off-diagonal WTDs, the second jump can occur immediately after the first, and thus $\lim_{\tau \to 0} w_{ss'}(\tau)$ for $s' \ne s$ is finite.  Clearly, this is not the case in the full dynamics and so we consider only the diagonal WTDs.
This difference between on- and off-diagonal WTDs only occurs at very short times and, as we will see in the next section, the ME description is unreliable in this regime, anyway.
\section{Numerical results}\label{results}
The main new finding obtained so far is our definition of the WTD via the continuous SE route. In the following we compare results obtained numerically from this definition with those from BD simulations (using the minimum-based jump definition), as well as with results from the ME approach.

To be specific, we choose the positions $A$, $B$, and $C$ to be adjacent minima of the potential $u(x)$ [see Eq.~(\ref{eq:pot})].
In the subsequent section~\ref{sec:results-sym}, 
we examine the case of zero drive ($F\!=\!0$) for two amplitudes of the periodic potential. Both symmetric and asymmetric initial (and boundary) 
conditions are considered.  In Sec.~\ref{sec:driven} 
we fix $u_0$ to $15\kBT$ and study driven systems for two values $F<F_c$ where $F_c=u_0\pi/a$ is the ``critical'' driving force beyond which the minima in $u(x)$ are eliminated.
\subsection{Zero drive ($F=0$)}\label{sec:results-sym}
We start by considering fully ``symmetric'' cases, where the potential is untilted ($F=0$), and the particle jumps from the centre of a minimum (say, $x=0$) to the right ($0\to a$) or to the left ($0\to -a$). 
Clearly, these jumps are characterised by the same WTD, i.e.,
$w_{-a}^{0\to a}(t)=w_{a}^{0\to -a}(t)$. Moreover, in such fully symmetric situations [with respect to the initial- and boundary conditions, and to $u(x)$], the WTD is proportional to the first passage distribution, i.e.~$w_{-a}^{0\to a}(t)=1/2\,F_{[-a,a]}(t;q_0^0)$ [see Eq.~(\ref{fptd})]. However, as we will demonstrate, the SE approach is more versatile when we deviate from the symmetric case.
We thus discuss the symmetric case as a reference.

\begin{figure}
	\centering
	\includegraphics[width=\linewidth]{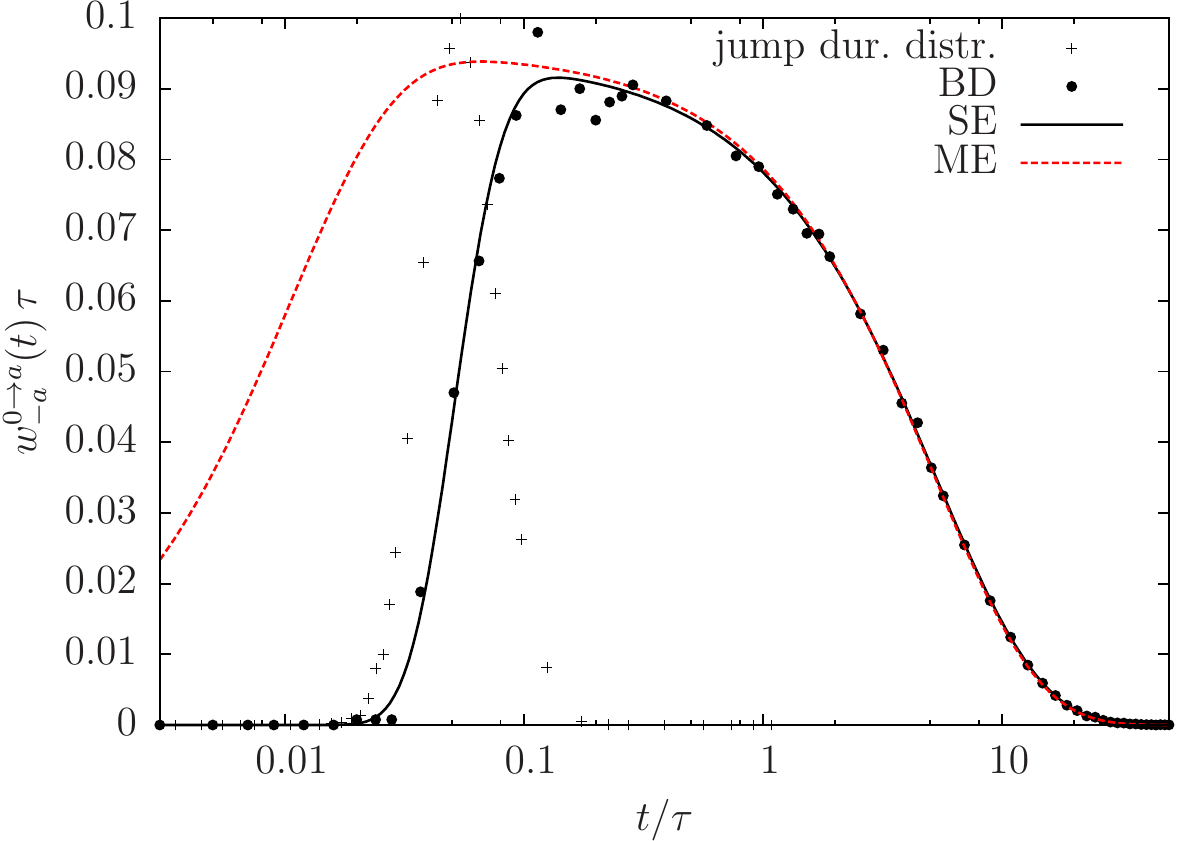}
	\caption{(Color online) Semilogarithmic plot of the WTD $w_{-a}^{0\to a}(t)$ for the symmetric case $F\!=\!0$, $u_0\!=\!5\kBT$ using all three methods BD, SE and ME (In the latter case, we have plotted $w_{++}(t)\!=\!w_{--}(t)$). The dashed line shows the jump duration distribution (from BD) with arbitrary units.}
	\label{wtdF0U5}
\end{figure}

\begin{figure}
	\centering
	\includegraphics[width=\linewidth]{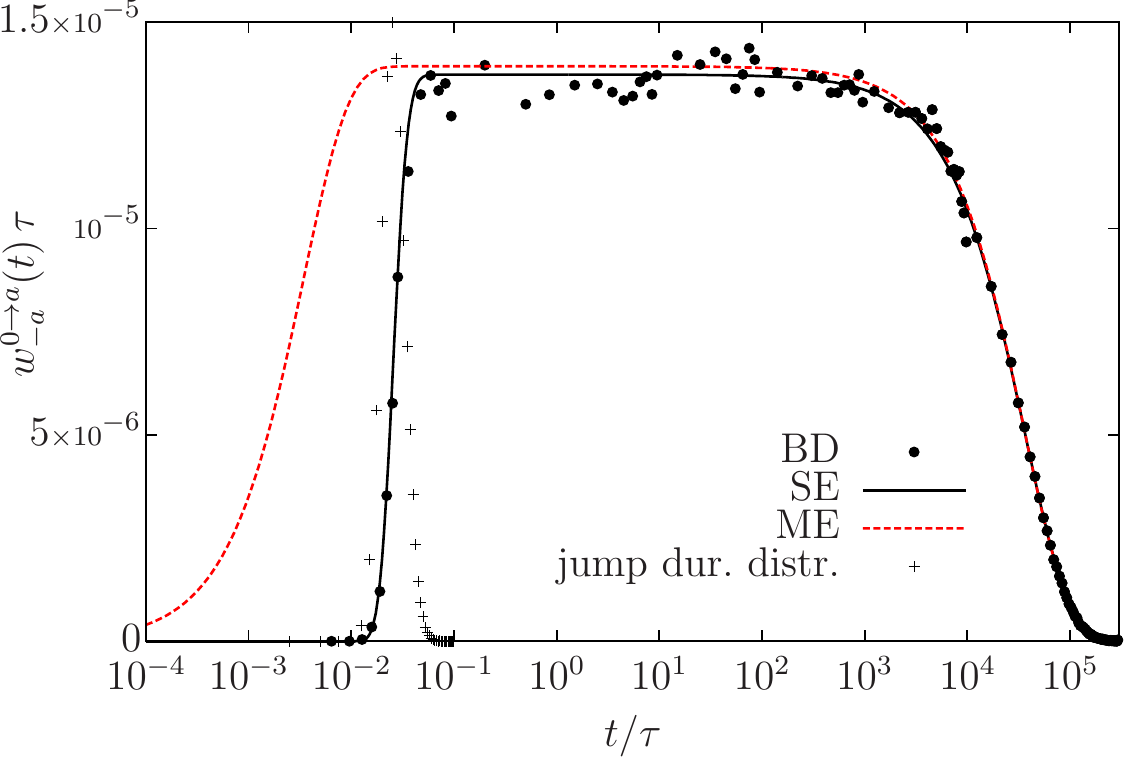}
	\caption{(Color online) Same as Fig.~\ref{wtdF0U5}, but for the ``deep well'' case, $u_0\!=\!15\kBT$.}
	\label{wtdF0U15}
\end{figure}

Results for the corresponding WTD at $u_0=5\kBT$ and $u_0=15\kBT$ are plotted in Figs.~\ref{wtdF0U5}
and \ref{wtdF0U15}, respectively.
These values of $u_0$ are representative since, as we recall from our earlier discussion of particle trajectories (see Figs.~\ref{trajectories} and \ref{trajectories-kramers}), the case $u_0=5\kBT$ is at the boundary between continuous and discrete motion while at $u_0=15\kBT$, the motion is essentially discrete. 

Figures~\ref{wtdF0U5} and \ref{wtdF0U15} contain data from all three approaches (SE, BD, ME). All curves share the same general structure in that 
the WTD is essentially (but not strictly) zero at very short times ($t\ll\tau$) and then grows with time up to a value which remains nearly constant over a range of intermediate times
(note the logarithmic time axis). Finally, at very large times the WTD decays rapidly to zero. The extension of the different regions and the actual values of the WTD depend on the potential height,
as we will analyse in detail below. From a methodological perspective we can state already here that the SE approach represents at all times a very accurate, 
smooth ``fit'' of the (somewhat noisy) BD data. The ME approach displays deviations at short times, to which we will come back at the end of this subsection.

The growth region of the WTD can be further analysed by inspection of the jump duration distribution obtained from BD, which is included in Figs.~\ref{wtdF0U5}
and \ref{wtdF0U15}. At both potential amplitudes this distribution displays a peak located at the (finite) time where the WTD grows most strongly. Moreover, the width 
of the peak corresponds approximately to that of the entire growth region. We can interpret these findings as follows.
First, the fact that the peak occurs at a {\em finite} time already signals that we are looking at (more or less) continuous stochastic processes where the particles needs a finite time
to cross a barrier (see Fig.~\ref{trajectories}(b)).  This is also the reason that we find a growth region in the WTD at all. As we have seen from the trajectories in Fig.~\ref{trajectories},
the jump time tends to decrease with increasing $u_0$. This explains the shift of the peak of the corresponding distribution towards earlier times in Fig.~\mbox{\ref{wtdF0U15}}.
Second, regarding the width of the jump duration distribution, we note that only realisations of the random force with a strong bias can push the particle against the potential ascent. The larger $u_0$, the smaller the fraction of appropriate noise realisations and hence the sharper the distribution. This also has a direct influence on the width of the growth region of the WTD: For very short times only very short jumps can contribute to the WTD, hence the WTD rises coincidently with the jump duration distribution.

\begin{figure}
	\centering
	\includegraphics[width=\linewidth]{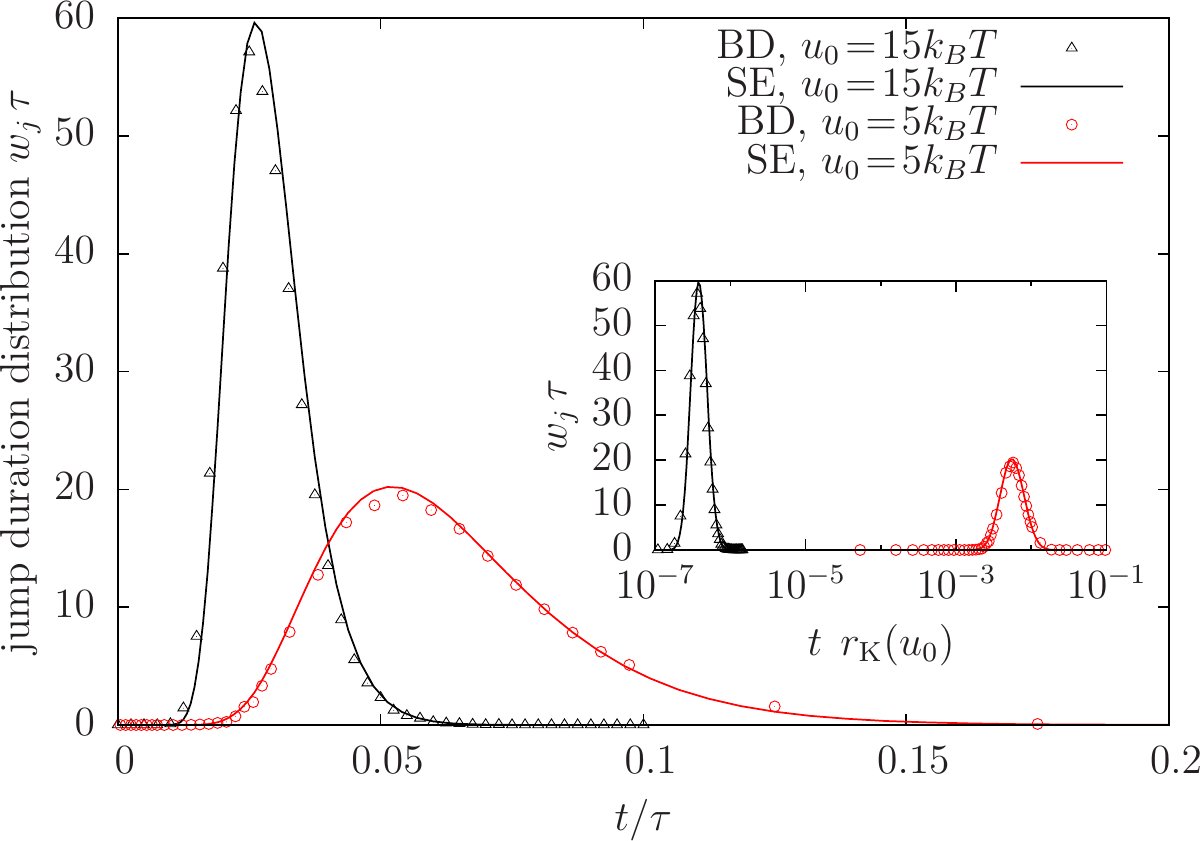}
	\caption{(Color online) Jump duration distribution $w_j(t)$ for $u_0\!=\!5\kBT$ and $u_0\!=\!15\kBT$ calculated with BD and SE [Eq.~\mbox{\eqref{jdSE}}]. The starting position is set to $\epsilon\!=\!0.01a$. Inset: Distributions $w_j(t)$ with the time axis being rescaled by Kramers' rate $r_\text{K}(u_0)$.}
	\label{fig:jd}
\end{figure}

Interestingly, it is also possible to calculate the jump duration distribution via the SE approach. 
To this end we just need to adjust the boundary and initial conditions. During a jump the particle leaves a minimum, say $x\!=\!0$, crosses the barrier and arrives at the next minimum, $x\!=\!a$. The difference to the jump $0\to a$ we considered above is that the particle actually leaves the minimum at $t\!=\!0$, i.e., it does not return. The essential step to the jump duration distribution is to consider an ensemble of realisations without processes where the particle comes back. Following Sec.~\mbox{\ref{sec:SE}} this condition corresponds to an absorbing boundary at $x\!=\!0$. The second absorbing boundary must be located at the point of arrival, $x\!=\!a$. This suggests the construction of the WTD $w_0^{\epsilon\to a}(t)$ where the starting position $\epsilon$ is close to $0$ because the particle just left the minimum.
Through the normalisation Eq.~\mbox{\eqref{normSE}} $w_0^{\epsilon\to a}(t)$ is connected with $w_a^{\epsilon\to 0}(t)$, which states how likely the particle comes back to $x\!=\!0$. We do not have any information about this quantity in the BD simulation, with which we compare. Therefore we need to remove this information. The total probability that the particle returns is given by $\int_0^\infty\mathrm{d}t\,w_a^{\epsilon\to 0}(t)$. This leads us to the following definition of the normalised jump duration distribution.
\begin{align}
	w_j(t)&=\frac{w_0^{\epsilon\to a}(t)}{1-\int_0^\infty\mathrm{d}t'\,w_a^{\epsilon\to 0}(t')}
	=
	\frac{w_0^{\epsilon\to a}(t)}{\int_0^\infty\mathrm{d}t'\,w_0^{\epsilon\to a}(t')}
	\label{jdSE}
\end{align}
A comparison between the BD and SE results for the distribution is given in Fig.~\ref{fig:jd}. Clearly, the SE route yields very good results.

At intermediate times the WTD has a broad maximum which, for deep wells, takes the form of a plateau (see Fig.~\ref{wtdF0U15}). 
Numerical values for the WTD maxima at the two potential amplitudes are given in Table~\ref{kramersrates}, where we have included values of Kramers' escape rate $r_\text{K}$ in the most precise form \cite{risken84}, i.e., without the Gaussian approximation of the integrals,
\begin{align}
	\label{Kramers}
	r_{\text{K},\infty}^{-1}=D_0^{-1}\int_{-a/2}^{a/2}\mathrm{d}x\,e^{-\beta u(x)}\int_0^{a}\mathrm{d}y\,e^{\beta u(y)}
	\,. 
\end{align}
This close relation can be understood on the basis of our SE approach where we have identified the WTD as a current
(i.e., a rate in one dimension). This current has been calculated with absorbing boundary conditions. A very similar calculation, namely by using one absorbing boundary and an infinite soft potential barrier on the other side together with the stationarity approximation \mbox{\cite{risken84,haenggi90}}, leads to Kramers' rate. Of course, for short and long times the probability distribution is not stationary; therefore the connection between the values of the WTD and Kramers' rate only holds at intermediate times.
\begin{table}[htbp]
	\centering
	\begin{tabular}{c|c|c|c}
		\hline\hline
		\parbox[c][0.4cm][b]{0.6cm}{$\displaystyle\!\frac{u_0}{\kBT}$} & global maximum & Kramers' rate \cite{risken84} & decay constant
		\\
		\hline
		$5$ &
		$0.0916\tau^{-1}$&
		$0.0924\tau^{-1}$ &		
		$0.1867\tau^{-1}$
		\\
		$15$ & 
		$1.39059 \ctimes 10^{-5}\tau^{-1}$&
		$1.39062 \ctimes 10^{-5}\tau^{-1}$& 
		$2.7953 \ctimes 10^{-5}\tau^{-1}$ 
		\\
		\hline
	\end{tabular}
	\caption{(Color online) SE results for the global maximum of the WTD $w_{-a}^{0\to a}(t)$, 
	Kramers' escape rate [see Eq.~(\ref{Kramers})], and the decay constant $\lambda$ characterising the exponential tail.}
	\label{kramersrates}
\end{table}

\begin{figure}
	\centering
	\includegraphics[width=\linewidth]{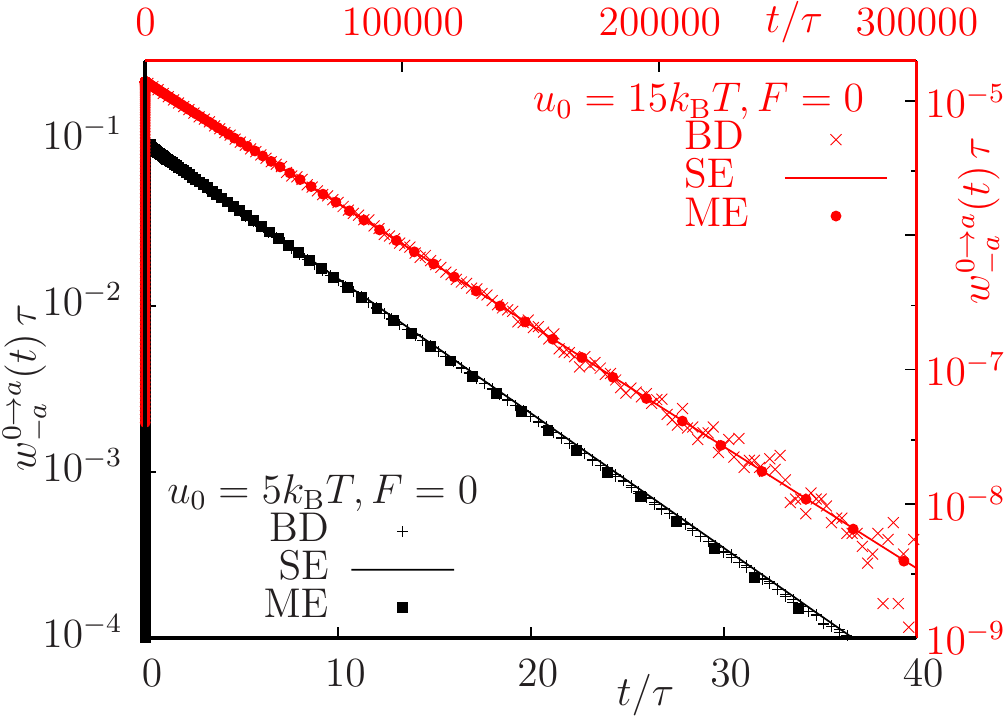}
	\caption{(Color online) Decay of $w_{-a}^{0\to a}(t)$ at long times for $u_0\!=\!5\kBT$ (left/bottom-axes) and $u_0\!=\!15\kBT$ (right/top-axes) for $F\!=\!0$ using all three methods (BD, SE, and ME). In the ME case we have plotted $w_{++}(t)\!=\!w_{--}(t)$. The decay constants are listed in Table \ref{kramersrates}.}
	\label{wtdF0-longtime}
\end{figure}

At times beyond the maximum (or plateau, respectively), the WTD rapidly decays to zero. In Fig.~\ref{wtdF0-longtime} we replot the corresponding behaviour with a linear time axis and a
logarithmic $y$-axis. From that it is clearly seen that the decay can be described by a (single) exponential, i.e., $w_{-a}^{0\to a}(t)\sim e^{-\lambda t}$ in this range of times. 
The corresponding decay constants $\lambda$ 
(as obtained from the SE approach) are listed in Table~\ref{kramersrates}. We find that, for both values of $u_0$ considered, $\lambda$ is approximately 
 twice as large as Kramers' escape rate (or plateau height, respectively). This is expected since the probability density is petering out simultaneously into both directions. Therefore the decay constant is expected to be the sum of Kramers' rates for every barrier, in the symmetric case $\lambda\approx 2r_\text{K}$.

We now turn to the ME approach. As described in Sec.~\ref{master}, this approach assumes a discretised
two-state-per-well model, that is, the particle can (only) take the positions $x_L$ and $x_R$ within each well. Leaving, e.g., position
$x_L$, the particle performs either an intra-well jump (to $x_R$) or an inter-well jump to $x_R-a$ or to $x_L+a$ (via $x_R$). Here we are only interested in processes of the latter type, where the particle actually crosses a barrier.  Further, because of translational symmetry and $F\!=\!0$ we only need to distinguish between the WTD for a ``long'' jump 
$w_\mathrm{long}(t)=w_{x_R-a}^{x_L\to x_L+a}(t)=w_{x_L+a}^{x_R\to x_R-a}(t)$, and the WTD for a ``short'' jump, $w_\mathrm{short}(t)=w_{x_R-a}^{x_R\to x_L+a}(t)=x_{x_L+a}^{x_L\to x_R-a}(t)$. 

The ME results shown in Figs.~\ref{wtdF0U5}, \ref{wtdF0U15}, and Fig.~\ref{wtdF0-longtime} pertain to a long jump. It is seen that the 
ME data become indeed consistent with those from the SE and BD approach, if one considers times {\em beyond} the growth region of the WTD.

Regarding the short-time behaviour, we find from Figs.~\ref{wtdF0U5} and \ref{wtdF0U15}
that the growth region of the WTD occurs at smaller times in the ME approach as compared to the SE and BD results. In other words, the WTD
predicts non-zero waiting times already below the mean jump duration predicted by BD. More precisely, the inflexion point of the WTD occurs at (roughly) the inverse
of the transition rate $\gamma\!=\!\gamma^++\gamma^-$ for {\em intra-well} transitions. In a linear approximation with respect to $x_R$ and $x_L$ of Eq.~(25) in \mbox{\cite{emary12}} one has $1/\gamma\!\approx\!\tau/(2u_0\pi^2)$,
which yields a good estimate of the inflexion point. We conclude that it is the intra-well relaxation 
(which, in turn, is governed by the potential amplitude) which is the main ingredient determining the short time behaviour of the WTD given by the ME. Indeed, because the ME is based on
a discrete model (and thus neglects travel times) we would not expect that the corresponding WTD is connected to the mean jump duration (as it was the case within the continuous SE approach).

\begin{figure}
	\centering
	\includegraphics[width=\linewidth]{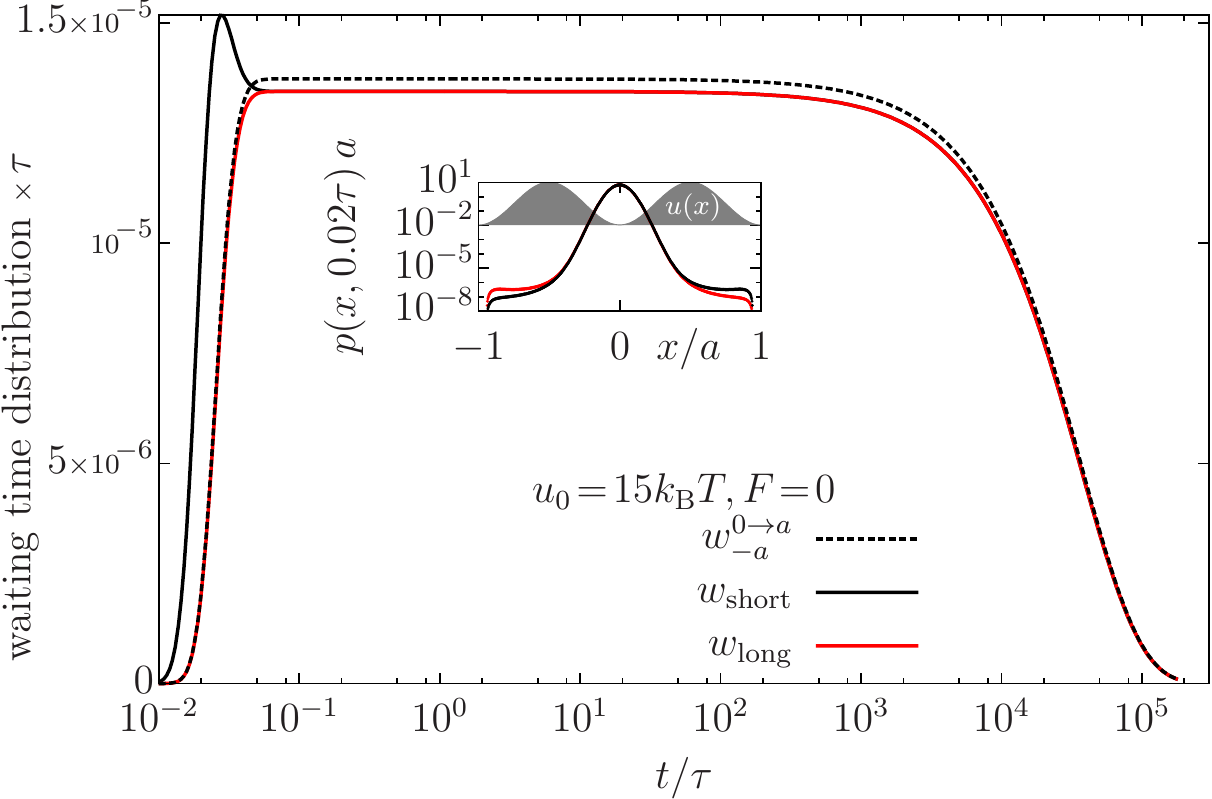}
	\caption{(Color online) Semilogarithmic plot of the WTDs $w_\mathrm{short}$, $w_\mathrm{long}$ and $w_{-a}^{0\to a}$ calculated with the SE%
	.  Inset: Probability density $p(x,t)$ at $t=0.02\tau$ for the long and short jump. The tiny plateaus at $\vert x\vert/a\approx 1$ determine the WTD.}
	\label{wtdF02sw}
\end{figure}

We finally note that asymmetric initial conditions such as those used in ME ($x_{L/R}\neq 0$)
can also be incorporated into the SE approach. As described before, we can distinguish between the WTD for a ``long'' jump, 
$w_\mathrm{long}(t)$, and the WTD for a ``short'' jump $w_\mathrm{short}(t)$.
In Fig.~\ref{wtdF02sw} we show these two WTDs together with our earlier SE result for $w_{-a}^{0\to a}(t)$. 
At short times, $w_\mathrm{short}(t)$ has a global maximum 
which exceeds the corresponding values of $w_\mathrm{long}(t)$ and the maximum of $w_{-a}^{0\to a}(t)$, that is, Kramers' escape rate. At subsequent time
the two WTDs $w_\mathrm{long/short}$ then merge at a value slightly below $w_{-a}^{0\to a}$. This is because the blip of $w_\mathrm{short}$ reduces the survival probability which in turn reduces the escape rate. The inset of Fig.~\mbox{\ref{wtdF02sw}} shows the probability densities $p(x,t)$ corresponding to $w_\mathrm{long}$ and $w_\mathrm{short}$ at $t\!=\!0.02\tau$ where the WTDs deviate from each other. At $t\!=\!0.02\tau$ the asymmetrically initialised densities have been broadened by diffusion to reach the boundaries, but yet not strong enough to eliminate the asymmetry.
\subsection{Driven system ($F>0$)}\label{sec:driven}
We now consider driven systems. Again, minima of the potential occur every $a$, the first being located at $x=m\colonequals a/(2\pi)\arcsin(F/F_c)$. Because of the tilt of the potential,
we have to distinguish between jumps to the left, $w^-(t)=w_{m+a}^{m\to m-a}(t)$, and jumps to the right, $w^+(t)=w_{m-a}^{m\to m+a}(t)$. Clearly, these WTDs provide
directional information, which cannot be extracted from the first passage time distribution.

\begin{figure}
	\centering
	\includegraphics[width=\linewidth]{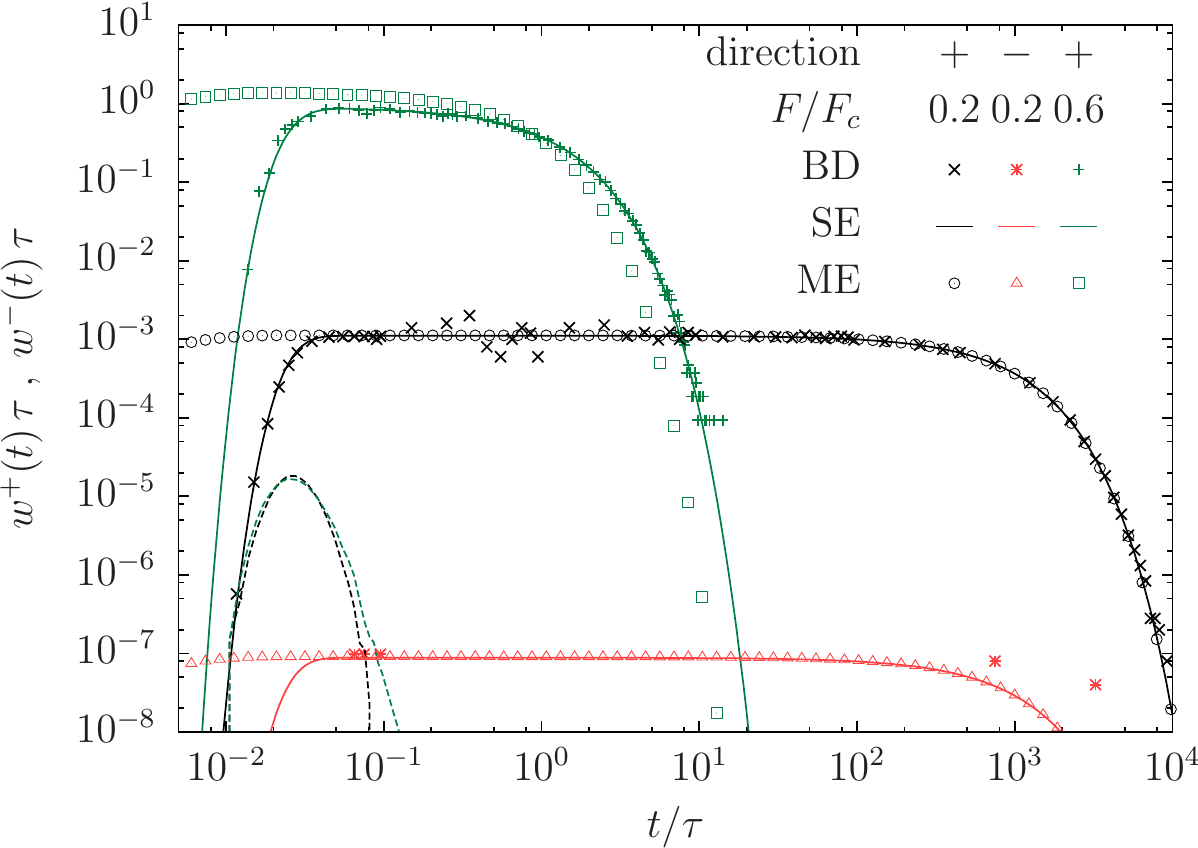}
	\caption{(Color online) Double-logarithmic plot of the WTD for the jump to the right $w^+(t)$ and to the left $w^-(t)$ (SE and BD) or $w_{++}(t)$ and $w_{--}(t)$ (ME), respectively. A positive force $F$ makes a jump to the left very unlikely which results in very small values for $w^-(t)$. Only for the smallest force $F=0.2F_c$ the BD recorded such events. Every red star is only a single jump.}
	\label{wtdFU15}
\end{figure}

Results for the WTDs in the ``deep well'' case are shown in Fig.~\ref{wtdFU15}. 
It is seen that the general structure of the WTD (consisting of a growth region, a plateau, and then a rapid decay) is the same as that at $F=0$ (see Fig.~\ref{wtdF0U15}). 
One main effect of increasing $F$ from zero is that the global maximum of $w^+$ increases, while the time range corresponding to the plateau shortens. In other words,
the average waiting time decreases and it occurs with larger probability. This is plausible, since a non-zero driving force leads to a lower {\em effective} barrier 
$\Delta u(u_0,F)=u_0(\sqrt{1-(F/F_c)^2}-F/F_c\arccos(F/F_c))$ in the driving direction. The opposite effects occurs against the driven direction, as reflected by the decreasing maxima
in $w^-(t)$. We also note that, for each $F$, the effects in the two WTDs are coupled via the the normalisation condition, i.e., the maximum in one of the WTDs can only grow
at the expense of the other one.

Comparing the different methods, we see from Fig.~\mbox{\ref{wtdFU15}} that the WTDs calculated with the SE agree with the BD results at all times. The WTDs given by the ME are consistent with SE and BD results for intermediate and long times. Again we observe a deviation for short times because, as discussed in the previous section, the short time scale in the two-state-per-well model describes intra-well relaxation and there is no time scale connected to the mean jump duration.
\section{Conclusion}\label{conclusion}
In present paper we have introduced and compared several routes to calculate the WTD in a system
with continuous, Markovian dynamics. Specifically, we have considered
the example of a Brownian particle in an one-dimensional tilted washboard potential, and focused on cases where the potential barriers are large
against the thermal energy. 

Traditionally, the WTD in such a situation is calculated by analysing trajectories obtained from BD simulations of the Langevin equation.
Here we define the WTD on the basis of the corresponding Smoluchowski equation (SE); i.e.,
we identify the WTD with one of the outgoing currents calculated from the SE with absorbing boundaries.
The resulting WTD is closely related to the first passage time distribution; however, the WTD contains directional information which is absent in the FPTD. This becomes particularly important in spatially asymmetric situations. Moreover, our definition of the WTD is more versatile than the FPTD in that it allows for unusual (asymmetric) initial and/or boundary conditions.  
Analysing a variety of systems with different initial conditions, as well as with and without external drive, we find in all cases full quantitative agreement
between the SE and BD results for the WTD. We also stress that, due to our rather general definition of the WTD in the SE approach, we are able
to calculate additional quantities such as the jump duration distribution. The latter is crucial for understanding the growth behaviour of the WTD at short times. 

In addition to the SE (and BD) method, we have also presented a fully analytical master equation approach to the WTD. The ME approach is based on the discrete,
two-state-per well model introduced by us in an earlier study \cite{emary12}. Comparing the results with those from SE and BD it turns out that the ME yields a very accurate 
WTD at intermediate times, where it exhibits a plateau, as well as at long times, where it has an exponential tail. Only the short-time behaviour differs due to the
fact that the ME model involves an ``intra-well'' relaxation time which is absent in the continuous approaches. 

From a computational point of view the BD route to the WTD seems, at first glance, to be the most straightforward one. However, closer inspection shows that the results
are quite dependent on the definition of ``jumps''. Furthermore, the results are often quite noisy. Here, the SE approach where noise is averaged out {\em a priori}, is clearly superior.
We stress, however, the solution is computationally expensive and ``fragile'' when using a standard solver. In Appendix~\ref{app:SE}
we have therefore sketched an alternative route to obtain the WTD from the SE based on a Fourier transform.

Finally, our SE approach to the WTD can be easily generalised to systems characterised by a different potential, to interacting systems,
and to systems with higher dimensionality. For example, for processes involving more than one spatial dimension, one would
simply replace the absorbing boundary value by an absorbing surface and calculate the WTD 
as an integral over the probability current over a part of the surface \cite{hartmann14}. Another generalisation concerns the character of the dynamics, which we here assumed to be Markovian.
For a non-Markovian SE (see \cite{metzler04,guill99,lichtner12,loos14} for examples) which involves memory kernels and thus has higher dimension in time, the FPTD can be applied straightforwardly \cite{metzler04,gang14,verechtchaguina06}. Therefore, and due to the similarity between WTD and FPTD we expect that
the identification of the WTD with the probability current remains intact. Finally, our generalisation of the WTD for continuous systems opens the route to calculate other quantities typically reserved for discrete (Markovian or non-Markovian) systems, such as the idle-time distribution \mbox{\cite{albert12}}. Work in these directions is in progress.
\begin{acknowledgments}
We gratefully acknowledge stimulating discussions with T.~Brandes.
This work was supported by the Deutsche Forschungsgemeinschaft
through SFB~910 (projects B2 and A7).
\end{acknowledgments}
\begin{appendix}
\section{WTD from BD simulations}
\label{app:BD}%
To obtain a reliable WTD the recorded histograms must have a high statistical quality at all times where the WTD is nonzero. Any deviation influences the normalisation and therefore shifts the entire WTD. Typically, the domain of non-zero WTD spans several orders of magnitude. Moreover, for large times the WTD decays exponentially (see Fig.~\ref{wtdF0-longtime}) which further enhances the computational effort due to the need of many long simulations. These considerations imply that the histogram bins must be sufficiently small to capture the time variations of the WTD and sufficiently large to yield a good statistical average. Here we use a piece-wise constant bin size distribution, which we adjust manually.
\par%
To achieve the necessary accuracy in the generated trajectories we choose a time step of $\Delta t=10^{-5}\tau$. This sets the mean size of a fluctuation to $4$\raisebox{.1\baselineskip}{$\scriptstyle\times$}$ 10^{-3}a$ which is necessary to sample the potential sufficiently accurate.
\par%
The simulation starts with the particle at the first minimum $m=a/(2\pi)\arcsin(F/F_c)$. During the simulation the trajectory is analysed to record leavings and arrivals of the particle at other minima $m\pm k a,\,k\in \mathbb{Z}$. Because one simulation alone cannot produce enough jumps in a reasonable computation time, we run several simulations in parallel. The least amount of jumps occur at $u_0\!=\!15\kBT$ and $F\!=\!0$. In this case we run $832$ simulations in parallel or successively, each computing $500$ jumps, to achieve a reasonable resolution for the long time part of the WTD.
\par%
Short waiting times occur even less frequently than large ones. To calculate the WTD at these short times a new set of simulations is started. Contrary to the regular trajectory simulation, the new simulations stop if the jump does not happen in a given time interval $u$ (with $\log_{10}(u/\tau) \in\{-1,\dots,2\}$). Each histogram for the Figs.~\mbox{\ref{wtdF0U5}, \ref{wtdF0U15} and \ref{wtdFU15}} is made of about $10^4$ of those short and $10^5$ long waiting times.
\section{WTD from the SE}
\label{app:SE}%
To calculate the WTD from the SE Eq.~\eqref{FP} we mainly use the FTCS (forward-time-centered-space) discretisation scheme \mbox{\cite{fletcher91}}. Because the whole process is determined by a very small current $j$ over the barrier and, correspondingly, small values for the probability density (see inset in Fig.~\mbox{\ref{wtdF02sw}}) we need very small discretisation steps $\Delta x$, a typical example being $\Delta x=0.0016a$. Nevertheless, the FTCS method works for all parameter sets except for $u_0\!=\!15\kBT$, $F\!=\!0$ where the solution of the SE just takes too long. To calculate the long-time part of the WTD in this particular case we use a different method involving a Fourier transform. To this end we write the probability density as
\begin{subequations}
\label{eq:ft}
\begin{align}
	p(x,t)&=\sum_{n=1}^N p_n(t)\,s_n(x)
	\,.
\end{align}
The sum involves the first $N$ modes
\begin{align}
	p_n(t)&=\frac{1}{a}\int_{-a}^{a}\mathrm{d}x\,s_n(x)\,p(x,t)
	\,,
\end{align}
where we exploit that for the special case $F\!=\!0$ with $A$, $B$, $C$ being successive minima of $u(x)$ the basis functions $s_n$ can be chosen as
\begin{align}
	s_n(x)&=\sin\left(\pi n\frac{x+a}{2a}\right)
	\,.
\end{align}
\end{subequations}
Applying the Fourier transform \mbox{\eqref{eq:ft}} to Eq.~\mbox{\eqref{FP}} yields
\begin{align}
	&\dot{p}_n(t)=\frac{\pi^2}{4a^2}\Big(-n^2 p_n
	\notag\\
	&+n U(p_{n-4}\Theta(n\!-\!4)-p_{n+4}-p_{4-n}\Theta(4\!-\!n))\Big)
	\label{fpft}
	\,,
\end{align}
where $\Theta(x)$ is the Heaviside-step function. The numerical solution of Eq.~\mbox{\eqref{fpft}} takes even longer than FTCS. Therefore we introduce the matrix based formulation $\dot{\vek{p}}(t)=A\cdot \vek{p}$
where $\vek{p}(t)$ contains all $p_n(t)$. The solution reads
\begin{gather}
	\vek{p}(t)=\exp(t A)\cdot\vek{p}(0)
	\label{vekpA}
	\,,
\end{gather}
which we calculate by decomposing the matrix $A$ into $S\cdot D\cdot S^{-1}$. In the latter expression the diagonal matrix $D$ consists of the eigenvalues of $A$ and $S$ contains the corresponding eigenvectors column by column. Inserting the diagonalisation in Eq.~\mbox{\eqref{vekpA}} we arrive at
\begin{align}
	\vek{p}(t)&=e^{t\,S\cdot D\cdot S^{-1}}\vek{p}(0)=S\cdot e^{tD}\cdot S^{-1}\cdot \vek{p}(0)
	\label{diagonalisation}
	\,,
\end{align}
where $e^{tD}$ is also diagonal. The computation of $\vek{p}(t)$ via Eq.~\mbox{\eqref{diagonalisation}} is fast and reliable. Specifically, for the case $u_0\!=\!15\kBT$ we find that $100$ Fourier modes are sufficient (computations finish in minutes even for $10^4$ modes).
\par%
A further advantage of the diagonalisation is that it provides a simple access to the long time dynamics. This is because the limit $\lim_{t\to\infty}\partial_t\ln w(t)$ gives the largest eigenvalue $d$
, and thus $w(t)\sim e^{dt}$. Note that $d$ is negative and close to zero.
\end{appendix}


\begin{thebibliography}{00}
%
\bibitem{montroll65} E.~W.~Montroll and G.~H.~Weiss, J.~Math.~Phys.~\textbf{6}, 167 (1965).
\bibitem{scher73} H.~Scher and M.~Lax, Phys.~Rev.~B \textbf{7}, 4491 (1973).
\bibitem{haus87} J.~W.~Haus and K.~W.~Kehr, Phys.~Rep.~\textbf{150}, 263 (1987).
\bibitem{higgins09} E.~R.~Higgins, H.~Schmidle, and M.~Falcke, J.~Theo.~Bio.~{\bf 259}, 338 (2009).
\bibitem{qian13} J.~Qian, Y.~Lin, H.~Jiang, and H.~Yao, Appl.~Phys.~Lett.~{\bf 103}, 223702 (2013).
%
%
\bibitem{saha12} S.~Saha, A.~Sinha and A.~Dua, J.~Chem.~Phys.~\textbf{137}, 045102 (2012).
\bibitem{english06} B.~P.~English, W.~Min, A.~M.~van Oijen, K.~T.~Lee, G.~Luo, H.~Sun, B.~J.~Cherayil, S.~C.~Kou, and
     X.~S.~Xie, Nat.~Chem.~Biol.~\textbf{2}, 87 (2006).
\bibitem{mainardi00} F.~Mainardi, M.~Raberto, R.~Gorenflo, and E.~Scalas, Physica A \textbf{287}, 468 (2000).
%
\bibitem{skaug13} M.~J.~Skaug, J.~Mabry, and D.~K.~Schwartz, Phys.~Rev.~Lett.~\textbf{110}, 256101 (2013).
\bibitem{prager06} T.~Prager and L.~Schimansky-Geier, J.~Stat.~Phys.~\textbf{123}, 391 (2006).
\bibitem{ganapathy10} R.~Ganapathy, M.~R.~Buckley, S.~J.~Gerbode, and I.~Cohen, Science \textbf{22}, 445 (2010).
\bibitem{hanes12} R.~D.~L.~Hanes, C.~Dalle-Ferrier, M.~Schmiedeberg, M.~C.~Jenkins, and S.~U.~Egelhaaf, Soft Matter \textbf{8}, 2714 (2012).
%
\bibitem{jung05} Y.~Jung, J.~P.~Garrahan, and D.~Chandler, J.~Chem.~Phys.~{\bf 123}, 084509 (2005).
\bibitem{chaud08} P.~Chaudhuri, Y.~Gao, L.~Berthier, M.~Kilfoil, and W.~Kob,
J.~Phys.: Condens.~Matter {\bf 20}, 244126 (2008).
\bibitem{ahn13} J.~W.~Ahn, B.~Falahee, C.~D.~Piccolo, M.~Vogel, and D.~Bingemann, J.~Chem.~Phys.~\textbf{138}, 12A527 (2013).
\bibitem{doliwa03} B.~Doliwa and A.~Heuer, Phys.~Rev.~E {\bf 67}, 030501(R).
\bibitem{heuer05} A.~Heuer, B.~Doliwa, and A.~Saksaengwijit, Phys.~Rev.~E {\bf 72}, 021503 (2005).
\bibitem{rubner08} O.~Rubner and A.~Heuer, Phys.~Rev.~E \textbf{78}, 011504 (2008).
%
\bibitem{hartmann14} C.~Hartmann, R.~Banisch, M.~Sarich, T.~Badowski, and C.~Schütte, Entropy \textbf{16}, 350 (2014).
\bibitem{thul07} R.~Thul and M.~Falcke, EPL \textbf{79}, 38008 (2007).
\bibitem{brandes08} T.~Brandes, Ann.~Phys.~(Berlin) \textbf{17}, 477 (2008).
\bibitem{flindt09} C.~Flindt, C.~Fricke, F.~Hohls, T.~Novotny, K.~Netocny, T.~Brandes, and R.~J.~Haug, Proc.~Nat.~Acad.~Sci.~\textbf{106}, 10116 (2009).
\bibitem{rajabi13} L.~Rajabi, C.~P\"oltl, and M.~Governale, Phys.~Rev.~Lett.~{\bf 111}, 067002 (2013).
\bibitem{albert11} M.~Albert, C.~Flindt, and M.~B\"uttiker, Phys.~Rev.~Lett.~\textbf{107}, 086805 (2011).
\bibitem{evers13} F.~Evers, R.~D.~L.~Hanes, C.~Zunke, R.~F.~Capellmann, J.~Bewerunge, C.~Dalle-Ferrier, M.~C.~Jenkins, I.~Ladadwa, A.~Heuer, R.~Casta\~neda-Priego, and S.~U.~Egelhaaf, Eur.~Phys.~J.~Spec.~Top.~\textbf{222}, 2995 (2013).
%
\bibitem{risken84} H.~Risken, {\it The Fokker-Planck Equation} (Springer, Berlin,
1984).
\bibitem{verechtchaguina06} T.~Verechtchaguina, I.~M.~Sokolov, and L. Schimansky-Geier, Europhys.~Lett.~\textbf{73}, 691 (2006).
\bibitem{benichou14} O.~B\'enichou, R.~Voituriez, Phys.~Rep.~\textbf{539}, 225 (2014).
\bibitem{metzler04} R.~Metzler and J.~Klafter, J.~Phys.~A: Math.~Gen.~\textbf{37}, R161 (2004).
\bibitem{haenggi90} P.~H\"anggi, P.~Talkner, M.~Borkovec, Rev.~Mod.~Phys.~\textbf{62}, 251 (1990).
\bibitem{gang14} G.~Gang , Q.~Xiaogang, Physica A \textbf{411}, 80 (2014).
\bibitem{shushin08} A.~I.~Shushin, Phys.~Rev.~E \textbf{77}, 031130 (2008).
\bibitem{sacerdote14} L.~Sacerdote, O.~Telve, and C.~Zucca, Adv.~Appl.~Prob.~\textbf{46}, 186 (2014).
\bibitem{navarro09} D.~J.~Navarro, I.~G.~Fuss, J.~Math.~Psychol.~\textbf{53}, 222 (2009).
\bibitem{Tierno10} P.~Tierno, P.~Reimann, T.~H.~Johansen, and F.~Sagues,
Phys.~Rev.~Lett.~{\bf 105}, 230602 (2010).
\bibitem{Dalle11} C.~Dalle-Ferrier, M.~Kr\"uger, R.~D.~L.~Hanes, S.~Walta, M.~C.~Jenkins, and S.~U.~Egelhaaf,
Soft Matter {\bf 7}, 2064 (2011).
\bibitem{xing05} J.~Xing, H.~Wang, and G.~Oster, Biophys.~J.~{\bf 89}, 1551 (2005).
\bibitem{Siler10} M.~Siler and P.~Zemanek,
New J.~Phys.~{\bf 12}, 083001 (2010).
\bibitem{Lopez08} B.~J.~Lopez, N.~J.~Kuwada, E.~M.~Craig, B.~R.~Long, and H.~Linke,
Phys.~Rev.~Lett.~{\bf 101}, 220601 (2008).
\bibitem{zapata96} I.~Zapata, R.~Bartussek, F.~Sols, and P.~H\"anggi, Phys.~Rev.~Lett.~\textbf{77}, 2292 (1996).
\bibitem{shah07} M.~A.~Shahzamanian, M.~Eatesami, and H.~Yavary, Supercond.~Sci.~Technol.~\textbf{20}, 640 (2007).
\bibitem{augello10} G.~Augello, D.~Valenti, B.~Spagnolo, Eur.~Phys.~J.~B \textbf{78}, 225 (2010).
\bibitem{bonnet11} I.~Bonnet and P.~Desboilles, Eur.~Phys.~J.~E \textbf{34}, 25 (2011).
\bibitem{kay07} E.~R.~Kay, D.~A.~Leigh, and F.~Zerbetto, Angew.~Chem.~Int.~Ed.~\textbf{46}, 72 (2007).
\bibitem{Ros05} A.~Ros, R.~Eichhorn, J.~Regtmeier, T.~T.~Duong, P.~Reimann, and D.~Anselmetti, 
Nature {\bf 436}, 928 (2005).
\bibitem{carinci13} G.~Carinci and S.~Luckhaus, J.~Stat.~Phys.~\textbf{151}, 870 (2013).
\bibitem{hairer08} M.~Hairer, G.~A.~Pavliotis, J.~Stat.~Phys.~\textbf{131}, 175 (2008).
\bibitem{emary12} C.~Emary, R.~Gernert, and S.~H.~L.~Klapp, Phys.~Rev.~E \textbf{86}, 061135 (2012).
\bibitem{Reimann01} P.~Reimann, C.~Van den Broeck, H.~Linke, P.~H\"anggi, J.~M.~Rubi, and A.~P\'erez-Madrid,
Phys.~Rev.~Lett.~{\bf 87}, 010602 (2001).
\bibitem{kloeden92} P.~E.~Kloeden, E.~Platen, \emph{Numerical Solution of Stochastic Differential Equations} (2nd edition, Springer, 1992).
\bibitem{ermak75} D.~L.~Ermak, J.~Chem.~Phys.~\textbf{62}, 4189 (1975).
\bibitem{kramers40} H.~A.~Kramers, Physica \textbf{7}, 284 (1940).
\bibitem{reimann99} P.~Reimann, G.~J.~Schmid, and P.~H\"anggi, Phys.~Rev.~E \textbf{60}, R1 (1999).
\bibitem{kramers4thorder} In this context, we have used the $4^{th}$ order approximation for Kramers' rate \mbox{\cite{risken84}}.
\bibitem{guill99} S.~Guillouzic, I.~L’Heureux, and A.~Longtin, Phys.~Rev.~E \textbf{59}, 3970 (1999).
\bibitem{lichtner12} K.~Lichtner, A.~Pototsky, and S.~H.~L.~Klapp, Phys.~Rev.~E \textbf{86}, 051405 (2012).
\bibitem{loos14} S.~A.~M.~Loos, R.~Gernert, S.~H.~L.~Klapp, Phys.~Rev.~E \textbf{89}, 052136 (2014).
\bibitem{albert12} M.~Albert, G.~Haack, C.~Flindt, M.~B\"uttiker, Phys.~Rev.~Lett.~\textbf{108}, 186806 (2012).
\bibitem{fletcher91} C.~A.~J.~Fletcher, \emph{Computational techniques for Fluid Dynamics, Volume 1} (2nd edition, Springer, 1991).
\end{thebibliography}
\end{document}